\documentclass[english]{ccdconf}

\usepackage{amsmath}
\usepackage{amssymb}
\usepackage{multirow}
\usepackage{caption}
\begin{document}

\title{Economic Dispatch of an Integrated Microgrid Based on the Dynamic Process of CCGT Plant}
\author{Zhiyi Lin, Chunyue Song*, Jun Zhao, Chao Yang, Huan Yin}

\affiliation{College of Control Science and Engineering,
Zhejiang University, Hangzhou, 310027
        \email{csong@zju.edu.cn}}

\maketitle


\begin{abstract}
Intra-day economic dispatch of an integrated microgrid is a fundamental requirement to integrate distributed generators. The dynamic energy flows in cogeneration units present challenges to the energy management of the microgrid. In this paper, a novel approximate dynamic programming (ADP) approach is proposed to solve this problem based on value function approximation, which is distinct with the consideration of the dynamic process constraints of the combined-cycle gas turbine (CCGT) plant. First, we mathematically formulate the multi-time periods decision problem as a finite-horizon Markov decision process. To deal with the thermodynamic process, an augmented state vector of CCGT is introduced. Second, the proposed VFA-ADP algorithm is employed to derive the near-optimal real-time operation strategies. In addition, to guarantee the monotonicity of piecewise linear function, we apply the SPAR algorithm in the update process. To validate the effectiveness of the proposed method, we conduct experiments with comparisons to some traditional optimization methods. The results indicate that our proposed ADP method achieves better performance on the economic dispatch of the microgrid.
\end{abstract}

\keywords{Microgrid, Dynamic Process, Combined-Cycle Gas Turbine, Approximate Dynamic Programming}

\footnotetext{This work was supported partially by the National Key Research and Development Program of China (No.2017YFA0700300), partially by the Key Research and Development Program of Guangdong (No.2020B0101050001), and partially by the Key Research and Development Program of Zhejiang Province (No.2021C01151).}

\section{INTRODUCTION}

In the past few decades, the increasing consumption of fossil energy has led to public interests and technical developments in utilizing various distributed energy sources. Distributed generations (DGs) with flexible operation modes have been proposed as a general strategy for improving energy efficiency, lowering curtailment, peak shaving, and shifting.

The DGs with high penetration make significant contributions to power variations, however, DGs also bring challenges to operate and maintain the stability of the power grid. To solve this problem, microgrids (MGs) have been viewed as an applicable solution to integrate various DGs, energy storage devices and loads, which are connected to the power grid as a whole controllable unit. In the area of energy, the economic dispatch (ED) of MGs is critical and has received extensive attention from research and industrial fields. 


In order to operate the MG safely and economically, several related studies have been proposed over the last decade. In \cite{bib1}, a dynamic programming-based algorithm was derived to solve the unit commitment problem in the MG,  including photovoltaic-based generators to reduce the economic cost. In \cite{bib3} and \cite{bib4}, day-ahead optimization for gas and power systems were studied, which also considered the partial differential constraints in natural gas transmission. 

For the models in system, though the aforementioned studies contribute considerably to the MG optimization problems, they did not consider the dynamic energy flow constraints of the cogeneration units. The neglect of dynamic process may result in false optimal solutions since the cogeneration unit is constrained by physical dynamic transitions. Therefore, we consider that a more precise model of MG is critical to obtain feasible solutions. In this paper, we construct a practical MG system consists of \emph{combined-cycle gas turbine (CCGT) plant}. Specifically, the CCGT plant is a typical cogeneration unit with a remarkable dynamic process, thus validating the effectiveness of the proposed method considering the dynamic energy flow constraints.

In this paper, we propose to use the intra-day optimization strategies, and solve the multi-time periods decision problem via dynamic programming (DP). However, the classical DP usually suffers from the "three curses of dimensionality" \cite{bib10} when handling high dimensional state space and action space. In this context, \emph{approximate dynamic programming (ADP)} is a promising real-time optimization method \cite{bib9}, which makes a trade-off between solvability and optimality of solutions. In the ADP framework, the large-scale optimization problem is viewed as \emph{Markov decision process (MDP)}, which is divided into small sub-problems and solved sequentially. The ADP algorithm has been demonstrated to be valid in resource allocation problems \cite{bib10} and energy storage management \cite{bib11}, while these works focused on the power systems. Therefore, how to apply ADP to operation MG with the CCGT plant still remains a problem.

To deal with this issue, this paper proposes a novel ADP algorithm for the economic dispatch of an integrated heat and power microgrid. Specifically, an autoregressive moving average (ARMA) multi-parameter identification model of the CCGT thermodynamic process is considered in the MG system model. We also design an ADP approach based on value function approximation (VFA), in which a post-decision state is employed to achieve the near-optimal solution to minimize the total operational cost over one day. Overall, compared to the existing works, the main contributions of this paper are listed as follows:
\begin{itemize}
	\item A finite-horizon MDP formulation is developed, which incorporates CCGT thermodynamic constraints. To keep the system Markovian, an augmented state vector of CCGT is introduced so that the principle of optimality holds. 
	\item A VFA-based ADP is proposed and achieves near-optimal solutions to the MDP model. As a result, the proposed method solves Bellman's equation forward iteratively.
	\item Numerical experiments on the proposed ADP method are conducted with comparisons to the traditional myopic policy and MPC policy, thus validating the effectiveness of the proposed ADP method.
\end{itemize}

The rest of this paper is organized as follows. Section 2 presents the MG system and the formulated MDP model in detail. Section 3 introduces the ADP solution and the VFA design. The experimental settings and comparisons are presented in Section 4. In Section 5, we draw conclusions and promising directions for the future.

\section{MODEL OF ECONOMIC DISPATCH FOR MICROGRID}

\begin{figure}
	\centering
	\includegraphics[width=\hsize]{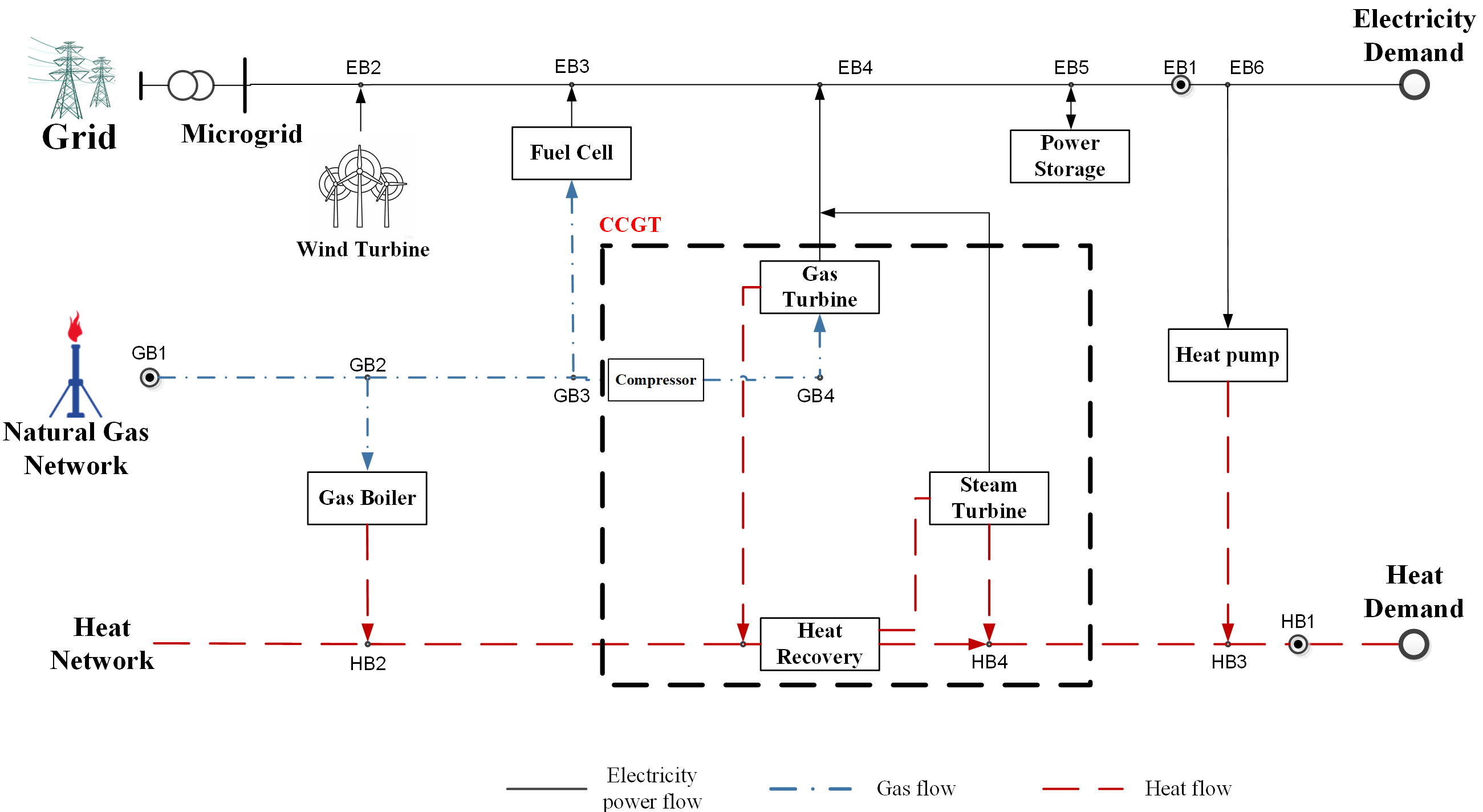}
	\caption{The schematic diagram of MG system.}
	\label{MG}
\end{figure}

This paper considers an integrated heat and power MG system, as shown in Figure~\ref{MG}, which consists of several dispatchable DGs: CCGT plant, gas boiler (GB), heat pump (HP), fuel cell (FC) and storage device. Wind turbines (WT) are renewable and non-dispatchable sources, which are also included in this MG system. We assume that the MG runs in a grid-connected mode and can trade with the upper-level grid according to the real-time electricity price. The heat and electricity demands on the users side are assembled in two load nodes respectively, which facilitates the information collection and comprehensive dispatch of the MG. Considering the intra-day operation of the integrated MG, we specify a finite time horizon $\mathcal{T}$, indexed by \{$\Delta t$, 2$\Delta t$,\ldots,$T$\}, where the time interval for each time step is $\Delta t$, and we set $\Delta t=15 min$, $T=96\Delta t=24h$ in a day. 

\subsection{The Markov Decision Process Formulation}
In the MG system, the real-time economic dispatch problem is a typical multi-time periods optimization problem, which can be decomposed into multiple sequential sub-problems and solved iteratively in the MDP framework. The basic elements of MDP are defined and introduced in this subsection.

The state variables are related to the minimally dimensioned function of MG, which are necessary to compute the decision function and the evolution of the system. The state vector $S_{t}$ at time step $t$ are defined in Equation~(\ref{eq1})-(\ref{eq3}), including both power system state variables $S_t^E$ and heat system state variables $S_t^H$ as follows:
\begin{equation}
\label{eq1}
S_{t}=  \left\lbrace S_t^E,S_t^H \right\rbrace 
\end{equation}
\begin{equation}
\begin{aligned}
\label{eq2}
S_t^E = \left\lbrace  P_{t-\Delta t}^{FC}, P_t^{CCGT}, SOC_t, P_t^{WT,a}, D_t^E, p_t \right\rbrace 
\end{aligned}
\end{equation}
\begin{equation}
\label{eq3}
S_t^H= \left\lbrace  Q_{t-\Delta t}^{GB}, Q_{t-\Delta t}^{HP}, \bar{Q}_t^{CCGT}, D_t^Q \right\rbrace
\end{equation}

where at time $t$, $D_t^E$ and $D_t^Q$ represent the electricity and heat demand of the system, $P_t^{CCGT}$ represents the active power output of CCGT, $SOC_t$ represents the state of charge of the power storage device, $P_t^{WT,a}$ represents the available wind power, $p_t$ represents the electricity price in real market, $\bar{Q}_t^{CCGT}$ represents the augmented states of CCGT thermal output $Q_t^{CCGT}$ (see Subsection\ref{2.2}). Some past system decisions in history are also included in $S_t$, e.g., $P_{t-\Delta t}^{FC}$, $Q_{t-\Delta t}^{GB}$ and $Q_{t-\Delta t}^{HP}$, since operational ramping constraints are considered in this paper. In this context, the feasible power output of units at $t$ is constrained by the previous states.

The decision variables at time $t$ include: the active power output of all dispatchable power generations, for example $P_t^{FC}$,$Q_t^{GB}$ and $Q_t^{HP}$; the natural gas input flow of CCGT $g_t^{CCGT}$; the charge and discharge state $u_t^{c}$,$u_t^{d}$ and power $P_t^c$,$P_t^d$; the active power of the MG exchanged with the upper-level grid $P_t^{Grid}$; and the curtailment power of WT $P_t^{wcur}$ and loads $P_t^{cur}$,$Q_t^{cur}$. This paper only focuses on active power balance of the electricity, since the MG system runs in the grid-connected mode which ensures node voltages and phase angles stability. Hence, the decision vector $x_t$ is described in Equation~(\ref{eq4}) as follows:
\begin{equation}
\begin{aligned}
\label{eq4} x_t  = \{ & P_t^{FC}, g_t^{CCGT}, P_t^{Grid}, P_t^c, P_t^d, u_t^{c}, u_t^{d}, \\ 
& P_t^{wcur}, P_t^{cur}, Q_t^{cur}, Q_t^{GB}, Q_t^{HP}, Q_t^{cur} \} 
\end{aligned}
\end{equation}

The exogenous information represents the stochastic factors in the system \cite{bib9}. In this paper, the exogenous information vector $W_t$ are the day-ahead forecast error of wind power generation $\hat P_t^{WT}$, real-time electricity price $\hat p_t$ and demands $\hat D_t^E$,$\hat D_t^Q$. $W_t$ is given by Equation~(\ref{eq5}) as follows:
\begin{equation}
\label{eq5} 
W_t  = \{ \hat P_t^{WT}, \hat D_t^E, \hat p_t, \hat D_t^Q\} 
\end{equation}

In the time sequence, the exogenous information $W_t$ arrives after the previous time step $t-\Delta t$ and before the current decision making at time $t$. Therefore, the decision process evolves as Equation~(\ref{eq6}).
\begin{equation}
\label{eq6} 
MG_t  = \{ S_0, x_0, W_{\Delta t}, \ldots, S_{t-\Delta t}, x_{t-\Delta t}, W_t, S_t\} 
\end{equation}

According to $S_t$, $x_t$ and $W_{t+\Delta t}$, the state transition function $S^M(S_t, x_t, W_{t+\Delta t})$ is determined by the following equations:
\begin{equation}
\label{eq7}
S_{t+\Delta t}^E(1) = P_t^{FC} 
\end{equation}
\begin{equation}
\label{eq8}
S_{t+\Delta t}^E(2) = a_0 + b_0\cdot g_t^{CCGT}
\end{equation}
\begin{equation}
\label{eq9}
S_{t+\Delta t}^E(3) = S_t^E(3)+(P_t^c\eta^c-\frac{P_t^d}{\eta^d})\cdot\Delta t
\end{equation}
\begin{equation}
\label{eq10} 
S_{t+\Delta t}^E(k)= P_{t+\Delta t}^F(k-3)+W_{t+\Delta t}(k-3),k\in\{4,5,6\}
\end{equation}
\begin{equation}
\label{eq11} 
S_{t+\Delta t}^H(1) = Q_t^{GB} ,\quad S_{t+\Delta t}^H(2)= Q_t^{HP}
\end{equation}
\begin{equation}
\label{eq12} 
S_{t+\Delta t}^H(3) = \boldsymbol A\cdot \bar Q_t^{CCGT} + \boldsymbol B\cdot g_t^{CCGT}
\end{equation}
\begin{equation}
\label{eq14} 
S_{t+\Delta t}^H(4) = P_{t+\Delta t}^F(4)+W_{t+\Delta t}(4)
\end{equation}

where  $P_{t+\Delta t}^F=\{ P_{t+\Delta t}^{WT,F}, D_{t+\Delta t}^{E,F}, p_{t+\Delta t}^F, D_{t+\Delta t}^{Q,F} \}$ represents the day-ahead forecast of exogenous information. Most units are modeled based on their popular energy hub model~\cite{bib10}, while the state transition function of CCGT is reformulated from the obtained ARMA model~(\ref{eq17}), which describes the dynamic process of CCGT. $\bar Q_t^{CCGT}$ represents the augmented states of the CCGT, and $\boldsymbol A$,$\boldsymbol B$ represents the coefficient matrices of $\bar Q_t^{CCGT}$ repectively.

The objective function $V_t^{*}(\cdot)$ is defined to minimize the total operation cost of the MG over the finite horizon $\mathcal{T}$. In time period $t$, the operation cost~(\ref{eq15}) is denoted by $C_t(\cdot)$, including fuel and operation cost $C_t^{f}(\cdot)$, cost of trading with grid $C_t^{tr}(\cdot)$, and penalties on the curtailment $C_t^{cur}(\cdot)$. Following Bellman's optimality principle, we design the optimal value funcition based on the state vector $S_{t}$, decision vector $x_t$ and exogenous information vector $W_t$ as follows: 
\begin{equation}
\label{eq15}
C_t(S_t,x_t)= C_t^{f}(S_t,x_t)+C_t^{tr}(S_t,x_t)+C_t^{cur}(S_t,x_t)
\end{equation}
\begin{equation}
\label{eq16}
\begin{aligned} 
V_t^{*} & =  \mathop{\min}_{x_t\in\mathcal{X}_t} \mathbb{E}\{\sum_{t=\Delta t}^T C_t(S_t, x_t) \} \\
& = \mathop{\min}_{x_t\in\mathcal{X}_t} (C_t(S_t, x_t)+\mathbb{E}[V_{t+\Delta t}(S_{t+\Delta t})|S_t,x_t])
\end{aligned}
\end{equation}
where $\mathcal{X}_t$ is the set of fesible decisions, and $\mathbb{E}(\cdot)$ is the conditional expectation.

\subsection{Dynamic Process of CCGT}\label{2.2}
The CCGT plant in the microgrid consists of the gas turbine, heat recovery system, steam turbine and corresponding controllers, etc. Obviously, the thermal power response of CCGT is slower than that of electric power due to the complex transient flow in the system. The consideration of this dynamic process makes our optimization work more reliable and distinct from the existing energy dispatch strategies for MGs. Some related work \cite{bib15} proposed an ARMA identification model considering the different response times of the CCGT plant. This paper transforms the ARMA model into high-order difference constraints, as shown in Equation~(\ref{eq17}), which are then integrated into the energy dispatch optimization model.

\begin{equation}
\label{eq17} Q^{CCGT}(k)\!=\!\sum_{m=1}^4 a_mQ^{CCGT}(k-m)+b_mg^{CCGT}(k-m-3)
\end{equation}
where $a_m$, $b_m$ are parameters estimated by means of system identification technique. $Q^{CCGT}(i)$ represents the heat output of CCGT at sampling point $i$, $g^{CCGT}(j)$ represents the natural gas flow input of CCGT at sampling point $j$. The sample interval is $50s$, thus there are 18 sampling points over one time period $t$. 

To make the decision process Markovian and have the prerequisite for applying DP, we reformulate the state variables in Equation~(\ref{eq3}) by adopting the augmented states $\bar{Q}^{CCGT}(k)$ as follows~\cite{bib16}: 
\begin{equation}
\label{eq18}
\bar{Q}_t^{CCGT}(k) = \left[
x_1(k)\quad x_2(k)\cdots x_6(k)\quad x_7(k)
\right]^{\mathrm{T}}
\end{equation}
\begin{small}
	\begin{equation}
	\begin{aligned}
	\label{eq19}
	\bar{Q}_t^{CCGT}(k+1) = \left[
	\begin{matrix}
	\boldsymbol 0 & \boldsymbol I\\
	0&\boldsymbol A_1
	\end{matrix}
	\right] 
	\!\cdot\! \bar{Q}^{CCGT}(k)+\left[
	\begin{matrix}
	\boldsymbol{0}^{\mathrm{T}}\quad 1
	\end{matrix}
	\right]^{\mathrm{T}}\!\cdot\! g_t^{CCGT}(k)
	\end{aligned}
	\end{equation}
\end{small}
\begin{equation}
\label{eq20}
Q_t^{CCGT}(k) = \left[
\begin{matrix}
b_4&b_3&b_2&b_1&0&0&0
\end{matrix}
\right]\bar{Q}^{CCGT}(k)
\end{equation}
where $k=1,2,\cdots,18$ in each time period $t$, $\boldsymbol I$ is $6 \! \times \! 6$ identity matrix, $\boldsymbol 0\!=\![0\quad0\quad0\quad0\quad0\quad0]^{\mathrm{T}}$, $\!\boldsymbol A_1=[0\quad 0\quad a_4 \quad a_3 \quad a_2 \quad a_1]$, respectively.

\subsection{Constraints}
In addition to the above thermodynamic constaints of CCGT, the objective function is subjected to the following constraints:
\begin{equation}
\begin{aligned}
\label{eq21}
& P_t^{FC}+P_t^{CCGT}+P_t^{Grid}+(P_t^{d}\cdot u_t^d-P_t^{c}\cdot u_t^c)\\
&+(P_t^{WT,a}-P_t^{wcur})-P_t^{HP}+P_t^{cur}=D_t^E
\end{aligned}
\end{equation}
\begin{equation}
\label{eq22}
Q_t^{GB}+Q_t^{CCGT}+Q_t^{HP}+Q_t^{cur}=D_t^Q
\end{equation}
\begin{equation}
\label{eq23}
\underline {P_t^i} \leq P_t^i \leq \overline{P_t^i}, i\in\{FC,CCGT,Grid\}
\end{equation}
\begin{equation}
\label{eq24}
\underline {Q_t^j} \leq Q_t^j \leq \overline{Q_t^j}, j\in\{GB,HP,CCGT\}
\end{equation}
\begin{equation}
\label{eq25}
R_t^{i,down}\cdot\Delta t \leq P_t^i-P_{t- \Delta t}^i \leq R_t^{i,up}\cdot\Delta t
\end{equation}
\begin{equation}
\label{eq26}
R_t^{j,down}\cdot\Delta t \leq Q_t^j-Q_{t- \Delta t}^j \leq R_t^{j,up}\cdot\Delta t
\end{equation}
\begin{equation}
\label{eq27}
u_t^{c}\cdot\underline {P_t^c} \leq P_t^c \leq u_t^{c}\cdot\overline {P_t^c}
\end{equation}
\begin{equation}
\label{eq28}
u_t^{d}\cdot\underline {P_t^d} \leq P_t^d \leq u_t^{d}\cdot\overline {P_t^d}
\end{equation}
\begin{equation}
\label{eq29}
u_t^{c}+u_t^{d}\leq1, u_t^{c},u_t^{d}\in\{0,1\}
\end{equation}
\begin{equation}
\label{eq30}
\underline{SOC}\leq SOC_t \leq \overline{SOC}
\end{equation}
\begin{equation}
\label{eq31}
0 \leq P_t^{wcur}\leq P_t^{WT,a}
\end{equation}
\begin{equation}
\label{eq32}
0 \leq P_t^{cur} \leq D_t^E, 0 \leq Q_t^{cur} \leq D_t^Q
\end{equation}
where Equation~(\ref{eq21}) and (\ref{eq22}) are the power and heat balance constraints of the MG respectively. The power generated from the dispatchable DGs and traded with the grid are limited by their lower and upper boundaries $\underline {P_t^i}$, $\overline{P_t^i}$, $\underline {Q_t^j}$, $\overline{Q_t^j}$, as indicated by Equation~(\ref{eq23})-(\ref{eq24}). Note that if $P_t^{Grid}$ is positive, the MG purchases electricity from the grid; otherwise, the MG sells surplus energy to the grid. The ramping rate of DGs is limited by Equation~(\ref{eq25})-(\ref{eq26}). The constraints of energy storage device are shown in Equation~(\ref{eq27})-(\ref{eq30}), while $u_t^{c}$ and $u_t^{d}$ are integer variables. The curtailment constraints for renewable power and demands are shown in Equation~(\ref{eq31})-(\ref{eq32}) respectively. All the aforementioned constraints in Equation~(\ref{eq18})-(\ref{eq32}), should be satisfied for time $t\in\mathcal{T}$.

\section{APPROXIMATE DYNAMIC PROGRAMMING SOLUTION}
In the framework of MDP, the typical multi-time periods decision problem can be solved recursively by dynamic programming. However, DP solves the Bellman's equation backward through time and explores every possible state at every time period, which constantly suffers from the curses of dimensionality for the large state and action space. To solve this problem, an improved alternative ADP is developed in this section. According to \cite{bib12}, ADP based on value function approximation (VFA) has been applied to obtain a near-optimal policy.
By approximating the value function around post-decision state variables $S_t^x$, the expectation form in Equation~(\ref{eq16}) is rephrased and the Bellman's equation is reformulated as a deterministic minimization problem as follows:
\begin{equation}
\label{eq34}
V_t(S_t)=\mathop{\min}_{x_t\in\mathcal{X}_t} (C_t(S_t, x_t)+\bar{V}_t^x(S_t^x)
\end{equation}
where $S_t^x$ is the state after the decision $x_t$ has been made but before the new exogenous information $W_{t+\Delta t}$ has arrived; $\bar{V}_t^x(S_t^x)$ is the VFA around $S_t^x$. Based on Equation~(\ref{eq34}), ADP is developed to solve the MDP problem forward at each time step. It is worth to note that the computation of the value function $V_t^x(S_t^x)=\mathbb{E}[V_{t+\Delta t}(S_{t+\Delta t})|S_t,x_t] $ is time-consuming and intractable. Therefore, a proper approximation $\bar{V}_t^x(S_t^x)$ to the optimal value function $V_t^x(S_t^x)$ is desired for guaranteeing the near-optimal policy $x_t^{*}$ according to current state information of the system. With this analysis, this paper proposes a piecewise linear function based ADP, and accomplishes the algorithm by learning the slopes of the optimal value function at the heat output state $Q_t^{CCGT,x}$.
\subsection{Piecewise Linear Function Approximation}
\begin{figure}
	\centering
	\includegraphics[height=5cm]{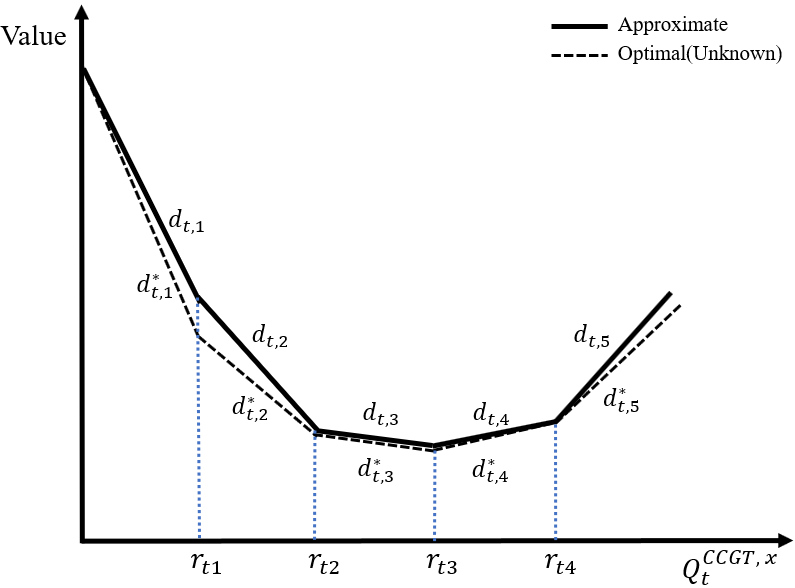}
	\caption{Optimal and approximate value function}
	\label{PLF_ADP}
\end{figure}

The approximate value function quantifies the long-term influence of the current decision $x_t$. In this paper, a convex piecewise linear function (PLF) is used to estimate the value of the heat output of CCGT according to \cite{bib13}, as presented by Equation~(\ref{eq35}).
\begin{equation}
\label{eq35}
\bar{V}_t^x(S_t^x)=\bar{V}_t^x(Q_t^{CCGT,x})=\sum_{a=1}^{N_t} d_{t,a}r_{t,a}, a\in\{1,\cdots,N_t\}
\end{equation}
where the slopes $d_{t,a}$ should be monotonically increasing, i.e., $d_{t,a}\leq d_{t,a+1}$. Actually, keeping convexity makes the optimization problem linear programs, which helps us handle the high-dimension state space and accelarates convergence. Figure~\ref{PLF_ADP} illustrates the exact optimal value function and the conducted approximation.
In time period $t$, the post-decision state $Q_t^{CCGT,x}$ equals to the heat output of CCGT at the last sampling point of this period, which can be calculated by the augmented state Equation~(\ref{eq18})-(\ref{eq20}). The post-decision state $Q_t^{CCGT,x}$ is then divided into $N_t$ segments on average, thus:
\begin{equation}
\label{eq36}
Q_t^{CCGT,x} = Q_t^{CCGT}(18)
\end{equation}
\begin{equation}
\label{eq37}
0\leq r_{t,a}\leq (\underline {Q_t^{CCGT}}-\overline{Q_t^{CCGT}})/N_t 
\end{equation}
\par We substitute~(\ref{eq35}) in the approximated Bellman's equation~(\ref{eq34}), then the near-optimal solution at time $t$ can be obtained by solving a deterministic optimization as follows:
\begin{equation}
\label{eq38}
x_t^{*} = \arg\mathop{\min}_{x_t\in\mathcal{X}_t,r_{t,a}\in\mathcal{R}_t} (C_t(S_t, x_t)+\sum_{a=1}^{N_t} d_{t,a}r_{t,a})
\end{equation}
where $\mathcal{R}_t$ is limited by Equation~(\ref{eq36}) and~(\ref{eq37}). Note that each time period is approximated by an independent PLF. 

\subsection{The Updating Process of PLF-ADP}
In order to make the decisions as close to the optimal as possible, the slopes of each segment for each time period should be updated iteratively until convergence. In this paper, we introduce the superscript $n$ to represent the variables value in the $n$th iteration. Therefore, $\bar{V}_t^{x,n-1}$ represents the approximate value function obtained in the $(n-1)$th iteration, which can be utilized to make decisions in the $n$th iteration as follows:
\begin{equation}
\begin{aligned}
\label{eq39}
V_t(S_t^n)&=\mathop{\min}_{x_t\in\mathcal{X}_t} (C_t(S_t^n, x_t^n)+\bar{V}_t^{x,n-1}(S_t^{x,n}))\\
&=\mathop{\min}_{x_t\in\mathcal{X}_t} (C_t(S_t^n, x_t^n)+\sum_{a=1}^{N_t} d_{t,a}^{n-1}r_{t,a}^n)
\end{aligned}
\end{equation}

To update the slopes of each segment, a sample observation of the marginal value $\hat{d}_{t-\Delta t,a}^n(Q_{t-\Delta t}^{CCGT,x,n})$ is needed, as indicated by Equation~(\ref{eq40}). 
\begin{equation}
\begin{aligned}
\label{eq40}
&\hat{d}_{t-\Delta t,a}^n(Q_{t-\Delta t}^{CCGT,x,n})=\hat{d}_{t,a}^n(Q_t^{CCGT,n})\\
&=V_t^{*}(Q_t^{CCGT,n})-V_t^{*}(Q_t^{CCGT,n}-\rho)
\end{aligned}
\end{equation}
Then the slopes of $\bar{V}_{t-\Delta t}^{x,n}(Q_{t-\Delta t}^{CCGT,x,n})$ can be updated as follows:
\begin{equation}
\begin{aligned}
\label{eq41}
&d_{t-\Delta t,a}^n(Q_{t-\Delta t}^{CCGT,x,n})=\alpha^{n-1}\hat{d}_{t,a}^n(Q_t^{CCGT,n})\\
&\qquad\qquad+(1-\alpha^{n-1})d_{t-\Delta t,a}^{n-1}(Q_{t-\Delta t}^{CCGT,x,n})
\end{aligned}
\end{equation}
where $\alpha^{n-1}$ is the stepsize to weight the information combined with the exsiting knowledge about the state value. There are several methods to decide the stepsizes, such as deterministic and stochastic stepsizes. In this work, a generalized harmonic stepsize rule is adopted to improve the rate of convergence. Note that Equation~(\ref{eq41}) only updates the slope for $a$th segment of $\bar{V}_{t-\Delta t}^{x,n}(Q_{t-\Delta t}^{CCGT,x,n})$. Besides, we apply the SPAR algorithm in \cite{bib9} to ensure the slopes are monotonically increasing after the update.

\begin{table}[!t]
	\renewcommand\arraystretch{1}
	\centering
	\caption{Parameters of Generators}
	\label{parameter1}
	\begin{tabular}{c|c|c|c|c}
		\hhline
		\multirow{2}*{Unit} & $P_{min}$ & $P_{max}$ & Ramp Rate & CC\\ 
		~ & (MW) & (MW) & (MW/h) & (\$/MWh) \\
		\hline
		FC         &	0.8	&	7	&	7	&	65\\ 
		\hline
		CCGT		&	6	&	43  &	38	&	92	\\
		\hline
		SOC        &	-3	&	3	&	-	&	-	\\ 
		\hline
		WT			&	0	&	3.6	&	-	&	-	\\
		\hline
		Grid		&	-6	&	6	&	6	&	$p_t$\\
		\hhline
	\end{tabular}
\end{table}

\begin{table}[!t]
	\centering
	\caption{Parameters of the Heat Generators}
	\label{parameter2}
	\begin{tabular}{c|c|c|c|c}
		\hhline
		\multirow{2}*{Unit} & $Q_{max}$ & $Q_{min}$ & Ramp Rate & CC\\ 
		~ & (MW) & (MW) & (MW/min) & (\$/MWh) \\
		\hline
		GB         &	1	&	15	&	3	&	300\\ 
		\hline
		HP     	&	0	&	5	&	5	&	-	\\ 
		\hline
		CCGT       &	15	&	50	&	0.5	&	-	\\ 
		\hhline
	\end{tabular}
\end{table}

\begin{table}[!t]
	\renewcommand\arraystretch{1}
	\centering
	\caption{Parameters of CCGT}
	\label{parameter3}
	\begin{tabular}{c|c|c|c|c}
		\hhline
		\multirow{4}{*}{Parameters} 
		& $a_1$ & $a_2$ & $a_3$ & $a_4$ \\ 
		\cline{2-5}
		~ &	1.6301	&	-0.6292	&	-0.3266	&	0.2570\\ 
		\cline{2-5}
		& $b_1$ & $b_2$ & $b_3$ & $b_4$	\\ 
		\cline{2-5}
		~ & 0.2087	&	0.06311	&	0.3656	&	0.4031	\\ 
		\hhline
	\end{tabular}
\end{table}

\section{Experiments}
In this section, the significance of considering the thermodynamic process of CCGT and the performance of the proposed PLF-ADP algorithm are validated by numerical experiments on an integrated heat and power microgrid system, as shown in Figure~\ref{MG}. The MDP model and associated constraints of the microgrid are available in Section 2. The parameters of the microgrid are partially shown in Table~\ref{parameter1}-\ref{parameter3}, where CC represents the cost coefficients of DGs in the optimization problem. The initial energy stored in the device is set to 7.5MW, meanwhile, the capacity and cycled efficiency of SOC are $\underline{SOC}=1.5$MW, $\overline{SOC}=15$MW, $\eta_t^c=\eta_t^d=0.9$. The penalties of curtailments $P_t^{wcur}$, $P_t^{cur}$ and $Q_t^{cur}$ are set to be 200\$/MWh, 150\$/MWh and 350\$/MWh respectively.

\begin{figure}[t]
	\centering
	\includegraphics[width=8cm]{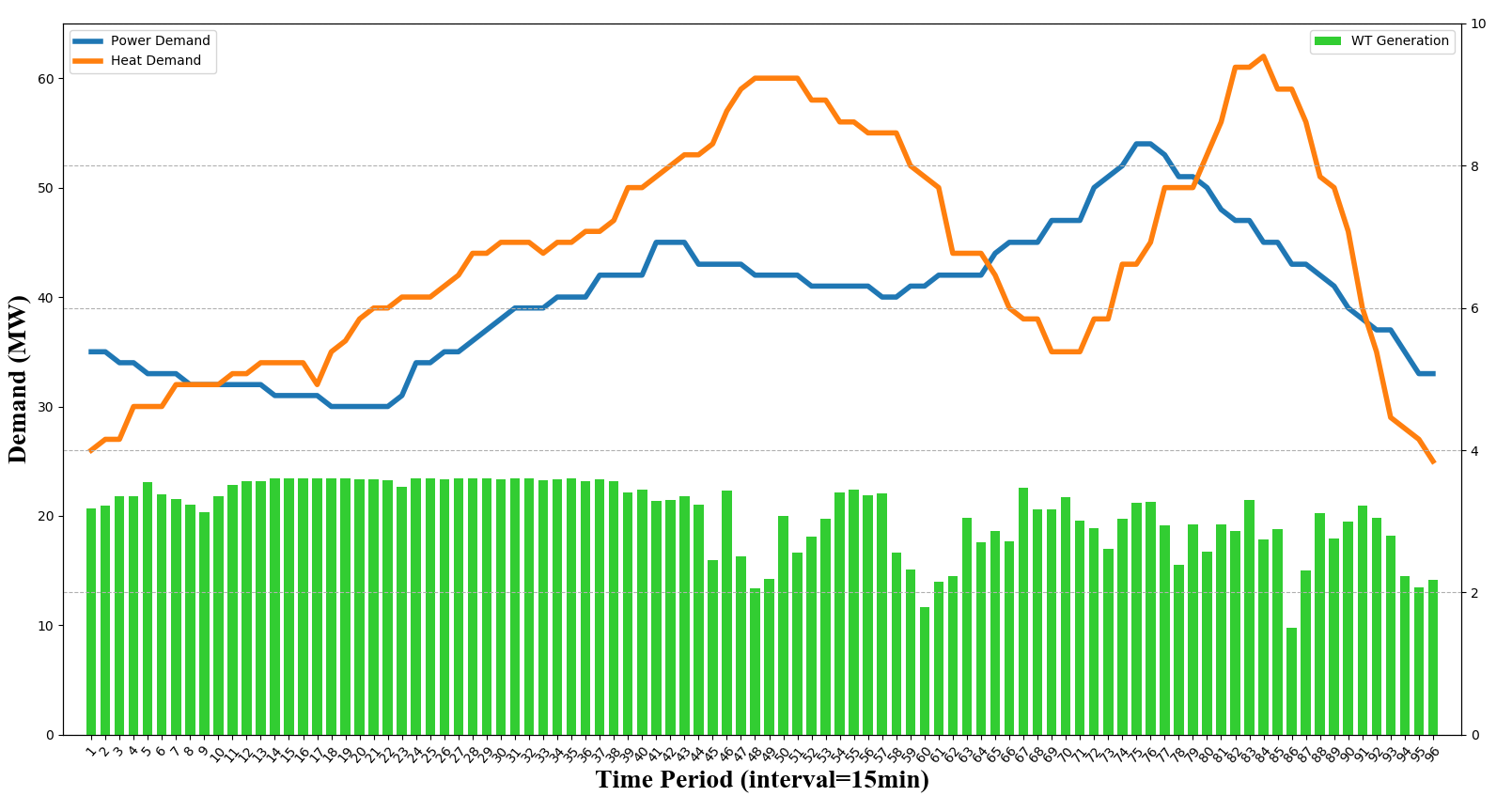}
	\caption{The Prediction of WT and Demands}
	\label{Demand}
\end{figure}
The day-ahead predicted power demand, heat demand and wind power are shown in Figure~\ref{Demand}. The wind power data in this paper comes from the real-world WT system in Turkey. The day-ahead prediction for electricity price of market is tiered pricing. All the experiments are conducted with Python on an Intel Core i5 2.80GHz Windows-based PC with 8GB RAM.

\begin{figure}[t]
	\centering
	\includegraphics[width=8cm]{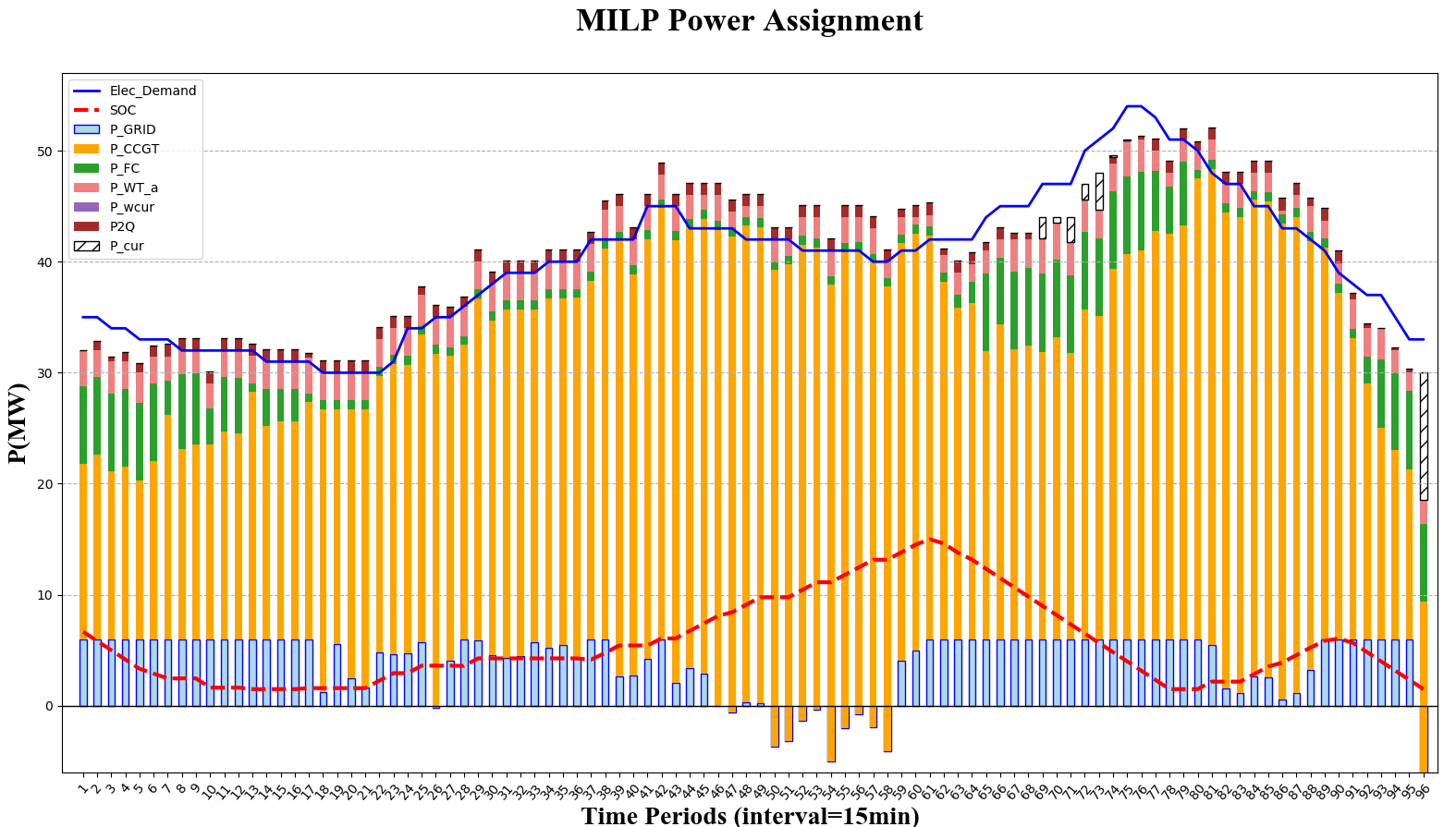}
	\caption{Day-ahead power dispatch based on MILP}
	\label{MILP_Power}
\end{figure}

\begin{figure}[!t]
	\centering
	\includegraphics[width=8cm]{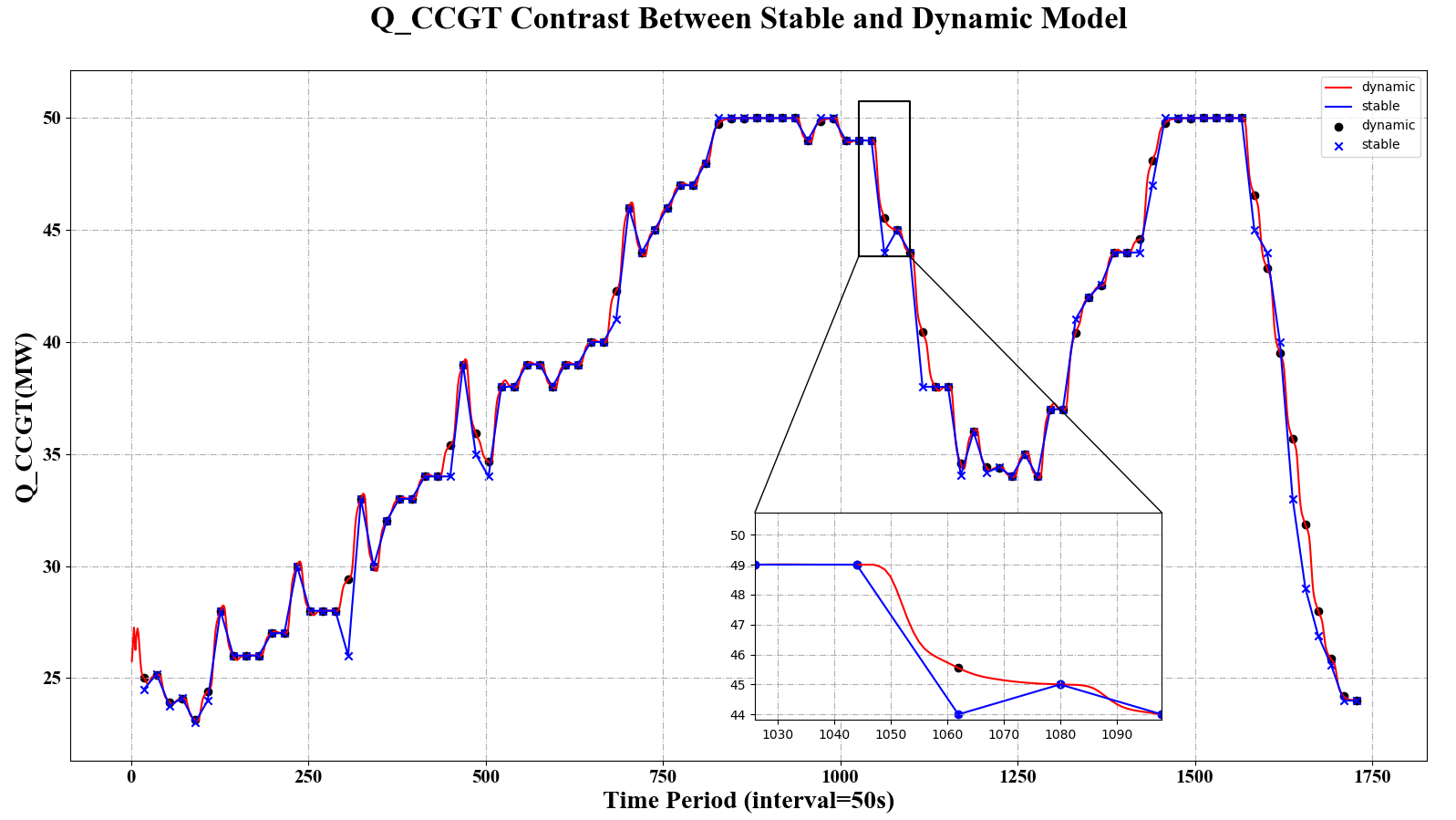}
	\caption{The Heat Output Curves of CCGT}
	\label{Stabel_Dynamic_Contrast}
\end{figure}

Firstly, the classical mixed integer linear programming (MILP) algorithm is implemented to obtain the operation strategy for the MG with the thermodynamic process of CCGT. The $\bar{Q}_t^{CCGT}$ augmented states model is used to constrain the feasible region of the input $g_t^{CCGT}$. Thus, the dynamic operation curve of the heat output $Q_t^{CCGT}(k)$ is recorded every 50 seconds, i.e., there are total 1728 sample points in 24h. 

The experimental results are shown in Figure~\ref{MILP_Power}-\ref{Stabel_Dynamic_Contrast}. It is obvious that the fluctuating electricity and heat demand is mainly provided by the CCGT and grid, while the potential thermal-electric coupling makes the auxiliary units necessary to satisfy Equation~(\ref{eq21})-(\ref{eq22}). After the midnight(0:00-6:00), the market electricity price is quite low and the MG tends to increase power purchase from the grid accordingly. Meanwhile, the energy storage device discharges to reduce the cost. In the time period 40-60(10:00-15:00), the demands gradually rise and the market price is relatively high, so the MG begins to sell power to the grid as much as possible on the basis of meeting the demands. Simultaneously, the power storage device is charged in advance to meet the load demand and reduce the load curtailment during load peak hours (time period 73-80). The gas boiler only generates heat in time periods 46-52 and 82-86 during which the heat output of CCGT almost reaches the upper limit. Figure~\ref{Stabel_Dynamic_Contrast} shows the heat output curves of the CCGT based on the energy hub model \cite{bib14} and identification model respectively, while the former model only considers the stable transition state of the CCGT. Figure~\ref{Stabel_Dynamic_Contrast} also shows that the two curves are not exactly the same and the curve with dynamic process is more smooth and practical for CCGT, since the stable curve may conflict with ramping constraints when the demands fluctuate quickly, thus demonstrating the significance of considering the thermodynamic process of CCGT.

\begin{figure}[t]
	\centering
	\includegraphics[width=8cm]{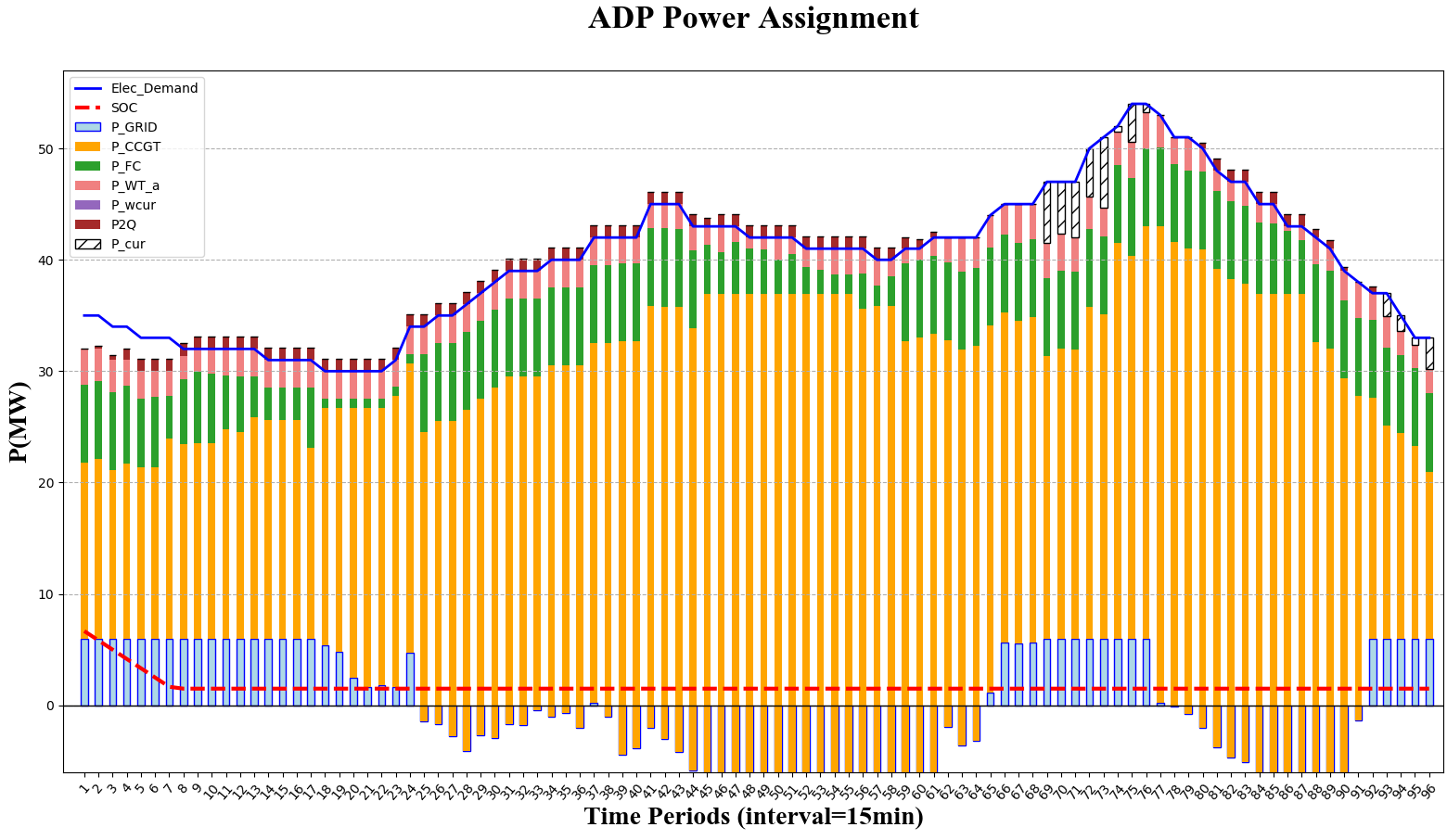}
	\caption{Intra-day power dispatch based on ADP}
	\label{ADP_Power}
\end{figure}

\begin{figure}[!t]
	\centering
	\includegraphics[width=8cm]{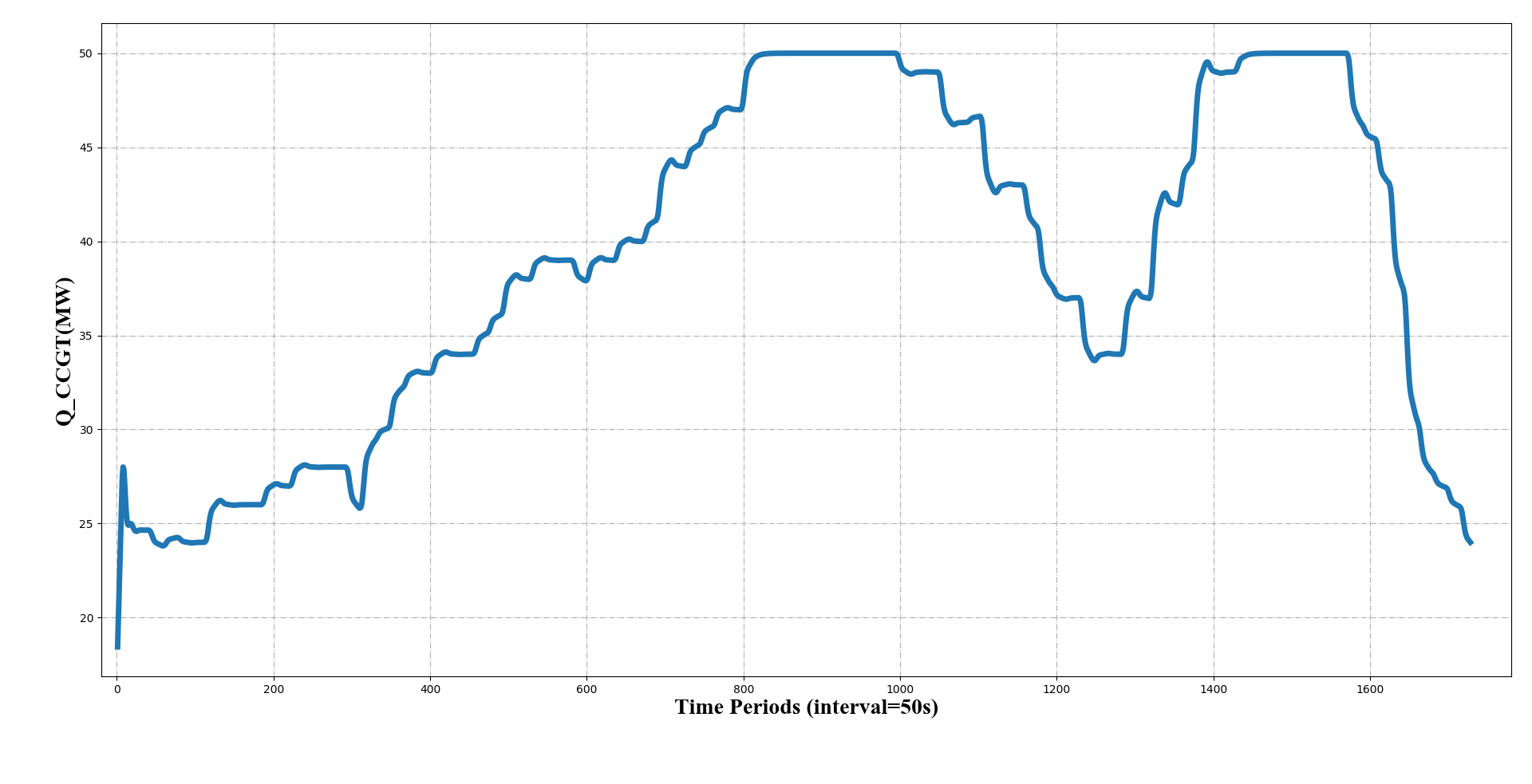}
	\caption{Heat output of CCGT based on ADP}
	\label{ADP_CCGT}
\end{figure}

\begin{figure}[!t]
	\centering
	\includegraphics[width=8cm]{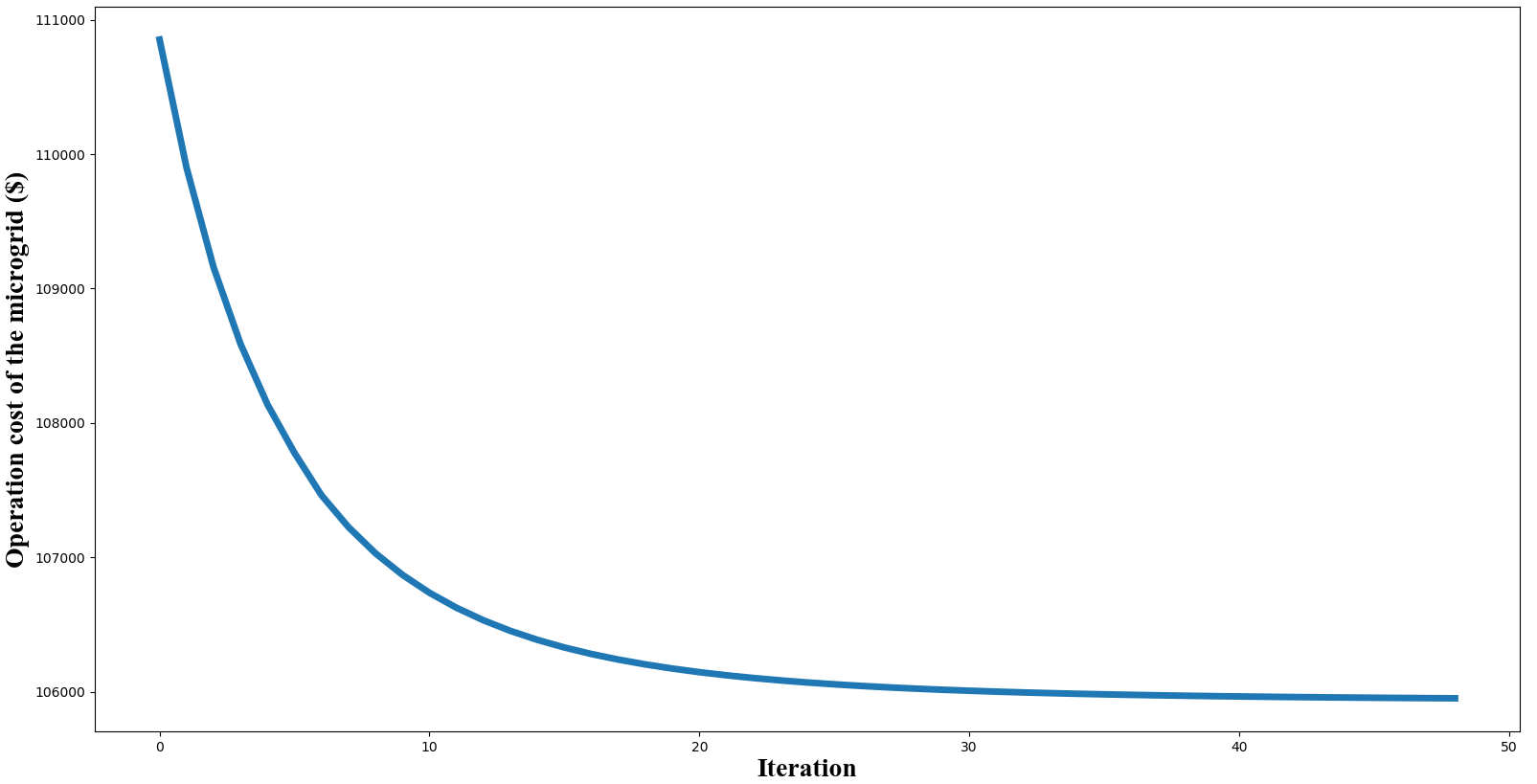}
	\caption{The convergence curve of ADP}
	\label{ADP_Iteration}
\end{figure}

Secondly, the performance of the proposed PLF-ADP algorithm is presented in Figure~\ref{ADP_Power}-\ref{ADP_Iteration}. The generation output of CCGT and the power exchange between the MG and the grid are shown in Figure~\ref{ADP_Power} and Figure~\ref{ADP_CCGT}. It is obvious that the MG sells electricity to the grid early in time period 24. The convergence process of the ADP is depicted in Figure~\ref{ADP_Iteration}, where the ADP converges in less than 40 iterations.

To demonstrate the effectiveness, myopic policy and model predictive control (MPC) are used as competitive comparisons. The experimental results show the solution from ADP performs better than the myopic policy and MPC algorithm and makes 5\% cost reduction, though the computation time is longer due to the iteration process. In summary, based on augmented states value approximation, the proposed PLF-ADP algorithm is effective for the economic dispatch of the integrated microgrid.

\section{CONCLUSIONS}
In this paper, we propose a novel ADP algorithm based on Markov decision process for the economic dispatch problem of a microgrid, which consists of both heat and power distributed generators in the real world. Specifically, we integrate the CCGT thermodynamic process into the approximate dynamic programming with augmented states. In the experimental section, we validate the effectiveness of the proposed algorithm with comparisons to the conventional optimization strategies. However, there remain some limitations in our work. Based on the existing research work, we intend to improve both performance and efficiency in the future, also with the uncertainty in the real applications, thus making the proposed ADP a more feasible and extensive application for automatic economic dispatch.

\balance

\end{document}